\documentclass[12pt,preprint]{aastex}

\newcommand{\teff}{$T_{\rm eff}$} 

\begin{document}

\title{Nitrogen abundances in giant stars of the globular cluster 
NGC 6752\footnote{Based on observations made with ESO Telescopes 
at the Paranal Observatories under programme 65.L-0165(A)}}

\author{David Yong}
\affil{Research School of Astronomy and Astrophysics, Australian National 
University, Mount Stromlo Observatory, Cotter Road, Weston Creek, ACT 2611, 
Australia}
\email{yong@mso.anu.edu.au}

\author{Frank Grundahl}
\affil{Department of Physics and Astronomy, University of Aarhus, Denmark}
\email{fgj@phys.au.dk}

\author{Jennifer A.\ Johnson}
\affil{Department of Astronomy, Ohio State University, Columbus, OH}
\email{jaj@astronomy.ohio-state.edu}

\author{Martin Asplund}
\affil{Max-Planck-Institut f{\" u}r Astrophysik, Karl-Schwarzschild-Str. 1, Postfach 1317, 85741 Garching b.\ M{\" u}nchen, Germany}
\email{asplund@mpa-garching.mpg.de}

\begin{abstract}

We present N abundances for 21 bright giants in the globular 
cluster NGC 6752 based on high-resolution UVES spectra of the 3360\AA~NH 
lines. We confirm that the Str{\" o}mgren $c_1$ index traces the N abundance 
and find that the star-to-star N abundance variation is 1.95 dex, at 
the sample's luminosity. We find statistically significant correlations, 
but small amplitude variations, 
between the abundances of N and $\alpha$-, Fe-peak, 
and $s$-process elements. Analyses using model 
atmospheres with appropriate N, O, Na, and Al abundances would strengthen, 
rather than mute, these correlations. If the small variations of heavy 
elements are real, then the synthesis of the N anomalies must take place 
in stars which also synthesize $\alpha$-, Fe-peak, and $s$-process elements. 
These correlations 
offer support for contributions from both AGB and massive stars 
to the globular cluster abundance anomalies. 

\end{abstract}

\keywords{Galaxy: Globular Clusters: Individual: NGC 6752, Galaxy: Stars: Abundances}

\section{Introduction}
\label{sec:intro}

Several decades have passed since the first identifications of 
the star-to-star abundance variations of light elements
in globular 
clusters \citep{popper47,osborn71,nf79}. 
Hydrogen burning via the CNO, Ne-Na, and Mg-Al chains 
at high temperatures qualitatively accounts for the 
observed trends: the abundances of C and O are low when N is high, 
O and Na are anticorrelated as are Mg and Al 
\citep{langer95,denissenkov98,karakas03}. 
While the amplitude of 
the variation may differ from cluster to cluster, the abundance patterns 
have been found in every well studied Galactic globular cluster 
\citep{smith87,kraft94,gratton04}. Such 
abundance patterns have also been identified in stars in  
extragalactic globular clusters \citep{letarte06,johnson06}. 
However, the origin of these abundance anomalies remains elusive. 

Internal nucleosynthesis and mixing can account for C, N, and Li in 
giant stars whose abundances exhibit a dependence upon evolutionary 
status \citep{ss91,grundahl02}.  
These abundance patterns, C and Li destruction along with N production, 
are also found in field stars as the evolve up the RGB \citep{gratton00b}. 
However, the identification of C, N, O, Na, Mg, and Al abundance 
variations in unevolved cluster stars 
\citep{briley96,gratton01,grundahl02,cohen05} 
demonstrates that an external pollution mechanism 
must be the dominant source since unevolved stars have insufficient 
internal temperatures to run the necessary nuclear reactions 
and lack a mechanism to mix the products to surface layers. 
Unlike C, N, and Li, abundance 
variations for the elements O to Al have rarely, if ever, been 
identified in field stars, indicating that the abundance anomalies 
are likely associated with some (presently unknown) property of the cluster 
environment. \citet{fulbright07} found low O and high Na and Al abundances 
in two bulge giants. However, they suggest that these two 
stars may be members of the bulge globular cluster NGC 6522. 
Intermediate-mass asymptotic giant branch (AGB) stars 
\citep{cottrell81,ventura05c} and 
massive stars \citep{prantzos06,smith06,decressin06}  
are candidates for the external 
pollution scenario. However, neither scenario 
currently provides a satisfactory explanation. 

Nitrogen abundance variations in globular cluster giants 
were inferred from early 
observations of CH and CN variations \citep{zinn77,dacosta80}. 
The NH molecular lines arguably 
offer the best probe of N abundances, in contrast to analyses 
of the CN lines which require knowledge of the C and O 
abundances. However, few studies have measured N from NH as well 
as the abundances for additional elements in a large 
homogeneous sample. In particular, few studies have measured N from NH 
using high-resolution spectra for a large sample. 
In this paper, we perform such an analysis of bright giants 
in the globular cluster NGC 6752 to measure the 
amplitude of the N abundance variation and to identify correlations 
between N abundances and light element abundances as well as possible 
correlations between N abundances and the abundances of heavy elements. 

\section{Observations and data reduction}
\label{sec:data}

\begin{figure}
\epsscale{0.8}
\plotone{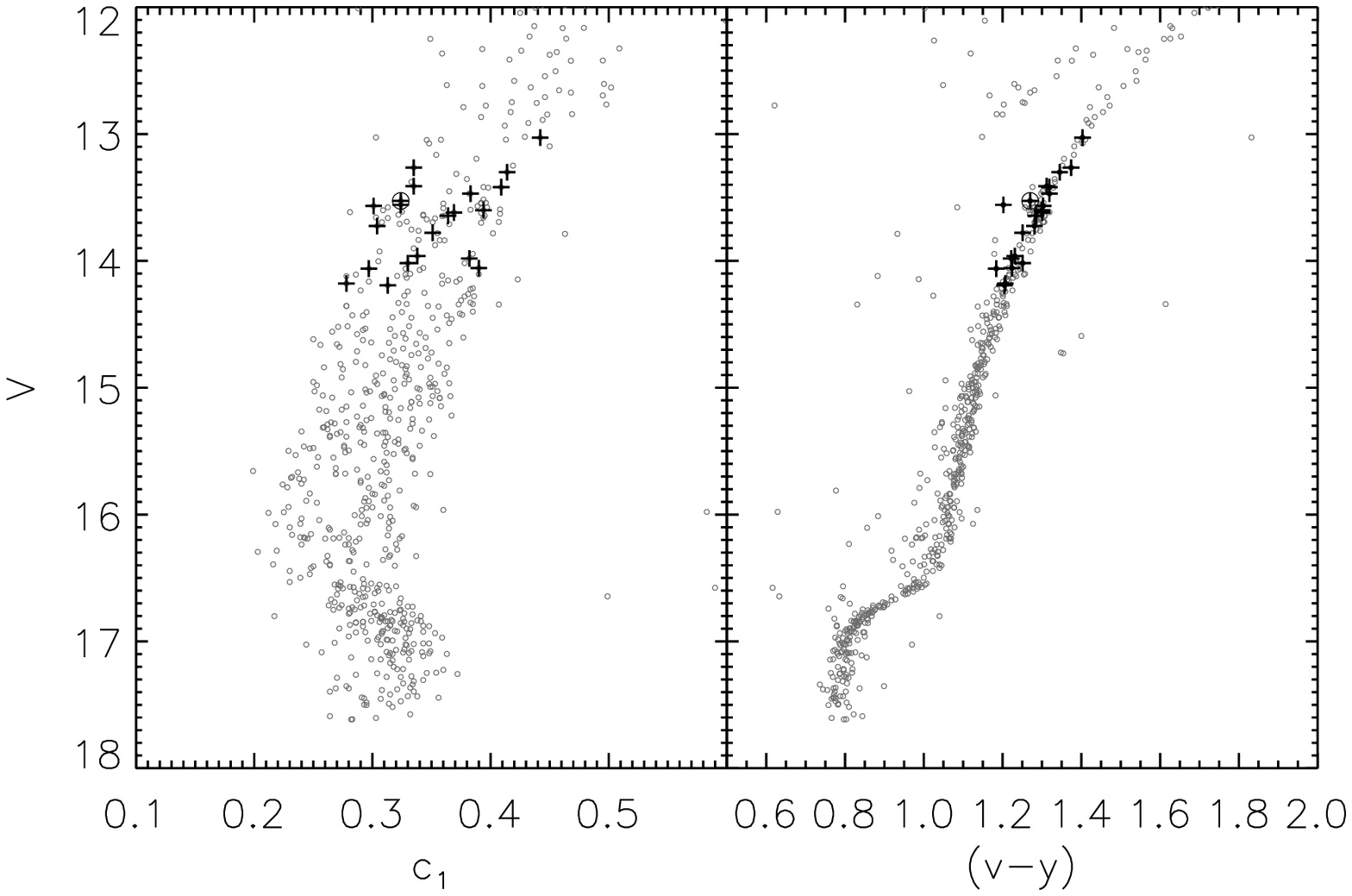}
\caption{The $V$ vs.\ $c_1$ (left) and $V$ vs.\ $(v-y)$ (right) 
color-magnitude diagrams using 
the \citet{grundahl99} photometry. The plus signs indicate the locations 
of our targets. (The large open circle marks NGC 6752-7.) \label{fig:cmd1}}
\end{figure}

The targets were the same as those analyzed by \citet{grundahl02}. 
The sample consists of 21 stars near the RGB bump, 
selected from the Str{\" o}mgren $uvby$ photometry of \citet{grundahl99}, 
and observed in service mode in Apr-May 2000 
using UVES \citep{uves}. 
The stars were selected with the expectation that they would sample 
the full range of the star-to-star light element abundance variation, 
at this luminosity. 
\citet{grundahl02} suspected that the Str{\" o}mgren $c_1$ index 
was tracing the N abundance (the $u$ filter includes 
the 3360\AA~NH features), and they 
identified clear correlations between $c_1$ and NH and CN indices 
defined by \citet{bs93}. 
The stars were deliberately selected 
to span the full range of the Str{\" o}mgren $c_1$ index.  
In Figure \ref{fig:cmd1}, we show the $V$ versus $(v-y)$ and the 
$V$ versus $c_1$ color-magnitude diagrams showing the locations of 
the targets. 

The ESO pipeline reduced spectra have a resolution of R = 60,000, but 
were smoothed with a 5 pixel boxcar function to increase the signal-to-noise
ratio (S/N). The resolving power after smoothing is R $\simeq$ 30,000. 
Although the S/N in the smoothed spectra is difficult to 
estimate in the crowded region near the 3360\AA~NH lines, 
we estimate that all stars have a
typical S/N = 70 per resolution element. For more details regarding
target selection, observations, and data reduction see \citet{grundahl02}. 

\section{Stellar parameters and abundance analysis}
\label{sec:analysis}

Derivation of the stellar parameters is described in \citet{grundahl02}. 
Briefly, the effective temperatures (\teff) were 
determined from the \citet{grundahl99} $uvby$ photometry using the 
\citet{alonso99b} color-temperature relations. The surface gravities 
($\log g$) were estimated using the stellar luminosities and \teff. 
The microturbulent velocities ($\xi_t$) 
were derived in the usual way by forcing 
the abundances from Fe\,{\sc i} to be independent of line strength. 
The stellar parameters for the program stars are presented in 
Table \ref{tab:param}. 

\begin{figure}
\epsscale{0.8}
\plotone{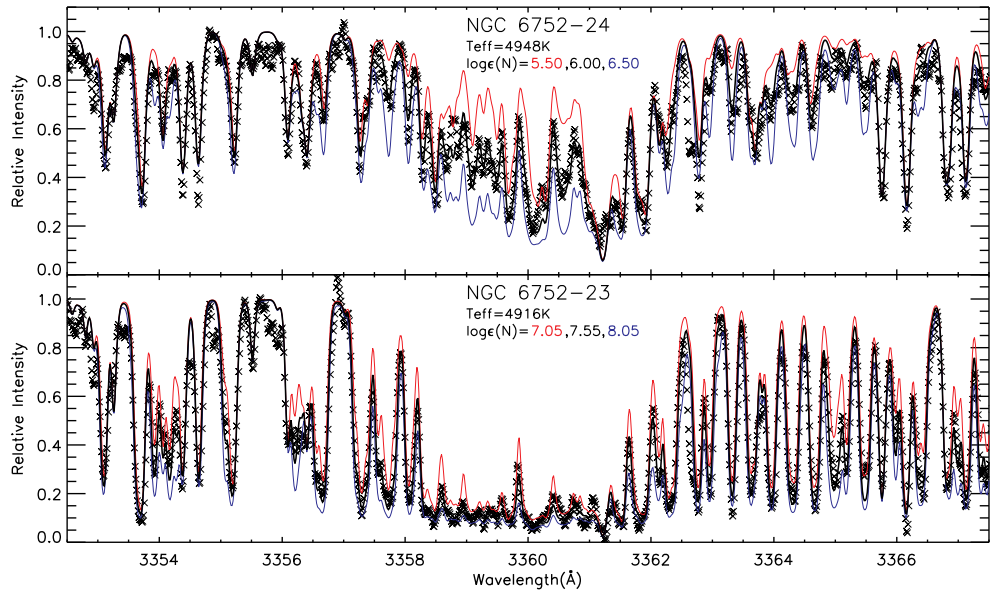}
\caption{Observed spectra (crosses) for NGC 6752-24 (upper panel) and 
NGC 6752-23 (lower panel) near the 3360\AA~NH band. 
These stars have very similar stellar parameters. 
Synthetic spectra with different N abundances are shown. 
The thick black line represents the best fit and unsatisfactory fits 
($\pm$ 0.5 dex) are shown as thin blue and thin red lines.\label{fig:1718fit}}
\end{figure}

\begin{figure}
\epsscale{0.8}
\plotone{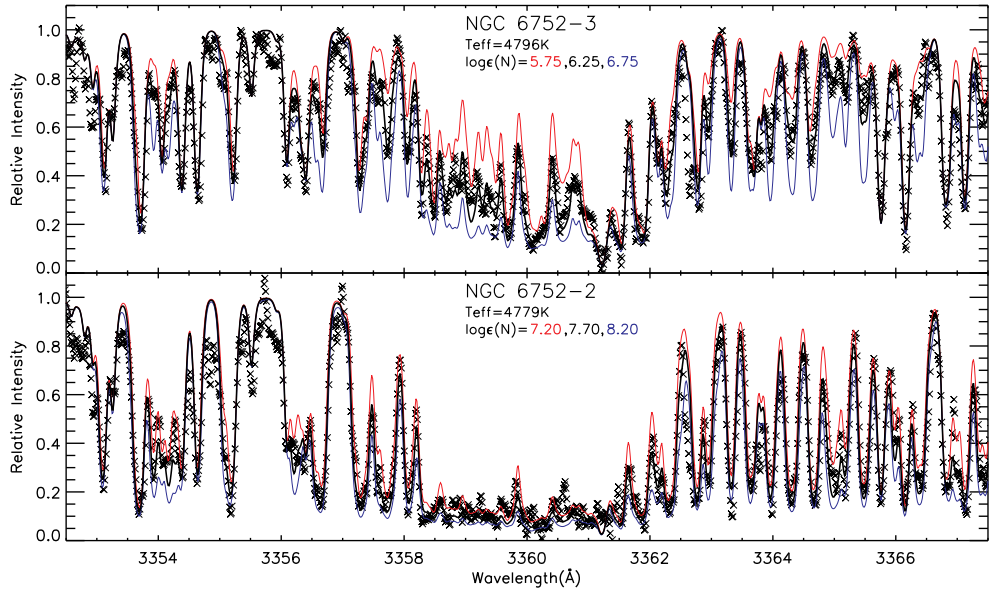}
\caption{Same as Figure \ref{fig:1718fit} but for the slightly 
cooler pair of stars 
NGC 6752-3 (upper) and NGC 6752-2 (lower).\label{fig:23fit}}
\end{figure}

Visual examination of the spectra and the considerable range in 
NH indices measured by \citet{grundahl02} 
indicate a large amplitude to the star-to-star 
N abundance variation. Quantitative N abundances were 
derived by comparing the observed spectra with
synthetic spectra generated using the LTE 
spectrum synthesis and line analysis package {\sc Moog} \citep{moog}. The 
molecular data and subsequent line list for the 
(0, 0) and (1, 1) bands of the 
$A-X$ electronic transition of the NH molecule at 3360\AA~were taken from 
\citet{johnson07}. We used model atmospheres computed by \citet{kurucz93}, 
which are the same as those used by \citet{6752,67522}. Synthetic spectra 
were generated with different N abundances (we assumed that all N was in 
the form $^{14}$N). The difficult task of setting the continuum was 
achieved by selecting a handful of continuum windows and 
using the synthetic spectra as a guide. We adjusted the N abundances 
until the residuals between 
the synthetic and observed spectra were minimized. 
In Figures \ref{fig:1718fit} and \ref{fig:23fit}, we show pairs of 
stars with essentially identical stellar parameters. However, these pairs 
of stars have considerably different N abundances as can be seen 
qualitatively by eye or quantitatively via the $c_1$ index and final 
[N/Fe] abundance. 
The N abundances, [N/Fe], are presented in Table \ref{tab:param} 
assuming the newly determined 
solar abundance of $\log$ $\epsilon$(N) = 7.78 \citep{grevesse07}. 
Our conclusions would not be changed had we adopted a previous value for 
the solar N abundance, e.g.,  
$\log$ $\epsilon$(N) = 7.92 from \citet{grevesse98}. 
 
We estimate that the internal uncertainties in the measured N abundances are 
typically 0.2 dex, resulting from errors in the adopted 
stellar parameters and errors in determining the best fit. 
The internal errors would be slightly lower ($\sim$0.15 dex) 
for the most N-poor stars and 
slightly higher ($\sim$0.25 dex) 
for the most N-rich stars due to saturation concerns. 
For additional elements discussed in 
Section \ref{sec:discussion}, 
the derivation of abundances and their associated uncertainties 
were described in \citet{grundahl02} and 
\citet{6752,67522}. (We omit Star NGC 6752-7 from further 
analysis due to its deviating [Fe/H].)

\section{Discussion}
\label{sec:discussion}

\subsection{Nitrogen abundances and comparisons with previous studies}

\begin{figure}
\epsscale{0.8}
\plotone{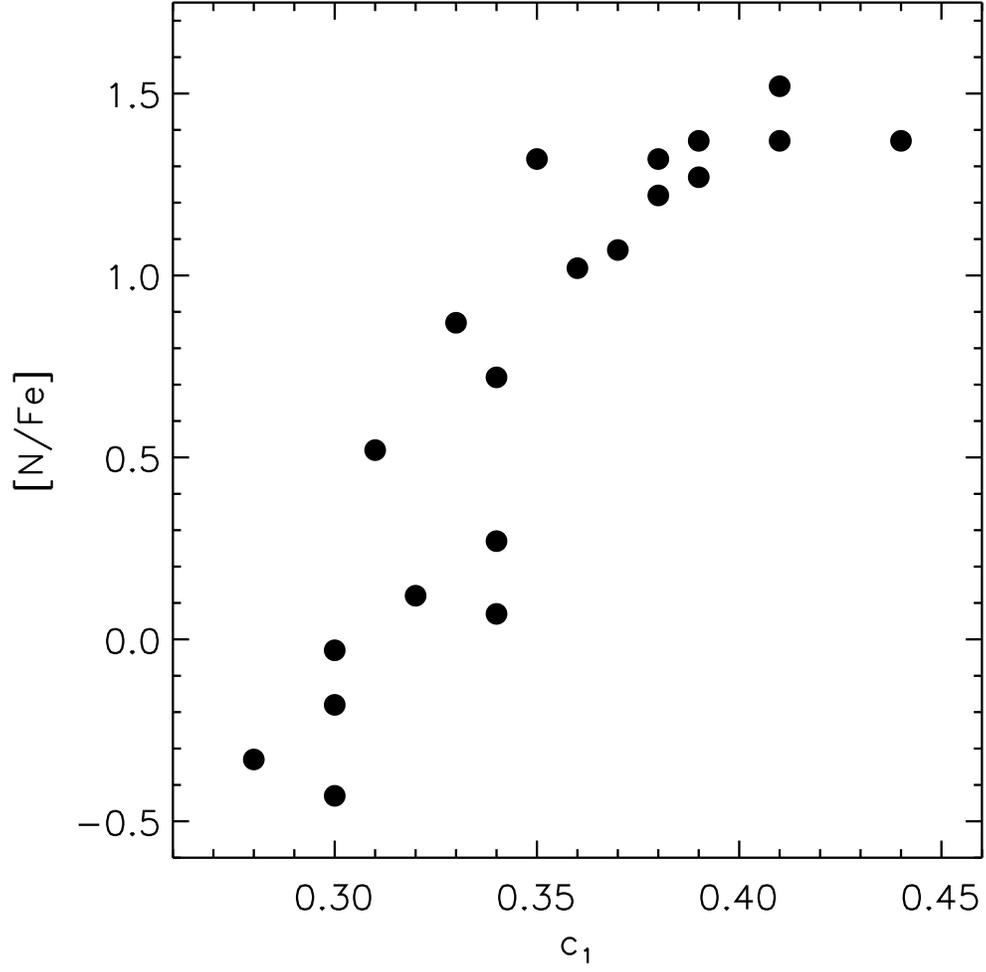}
\caption{[N/Fe] vs.\ $c_1$. Star NGC 6752-7 is not included 
in this plot due to its deviating [Fe/H]. \label{fig:nc1}}
\end{figure}

The [N/Fe] ratios for the sample range from [N/Fe] = $-$0.43 to 
[N/Fe] = +1.52, almost a 2 dex variation. 
Recall that the 
sample was selected according to the Str{\" o}mgren $c_1$ index, which 
\citet{grundahl02} suspected was tracing the 
N abundance. In Figure \ref{fig:nc1}, we plot the 
$c_1$ index versus the N abundance. The clear correlation 
confirms that $c_1$ indeed traces the N abundance. Given the 
Str{\" o}mgren $c_1$ selection criteria, it is not surprising 
that the full abundance distribution is well sampled. However, 
our N abundances are not representative of the N distribution for 
the entire cluster. 

\citet{carretta05} measured N abundances for 9 dwarfs and 9 subgiants in 
NGC 6752. Their N measurements are based on analysis of 
the CN molecular lines (rather than the NH lines used in this analysis) 
which requires knowledge of the C and O abundances.  
Our syntheses of the NH lines, and therefore our N abundances, 
are entirely independent of the 
adopted C and O abundances and presumably the uncertainties in 
our N abundances will be lower than analyses based on CN features. 
Aside from 1 subgiant with the abundance ratio [N/Fe] = 0.0, 
the \citet{carretta05} sample spans the range +1.0 $\le$ [N/Fe] $\le$ +1.7. 
Their maximum abundance [N/Fe] = +1.7 is comparable to the maximum 
abundance derived in this study, within the measurement uncertainties. 
However, their lowest abundance [N/Fe] = 0.0 is considerably higher 
than the minimum value found in our analysis. If CN cycling 
has occurred in our stars, which are more evolved than the
\citet{carretta05} sample, we would expect our stars to have 
systematically higher N abundances (assuming 
that the two samples were drawn from 
the same population and that the two samples covered the full range of the 
abundance variation at their respective luminosities). 

\begin{figure}
\epsscale{0.8}
\plotone{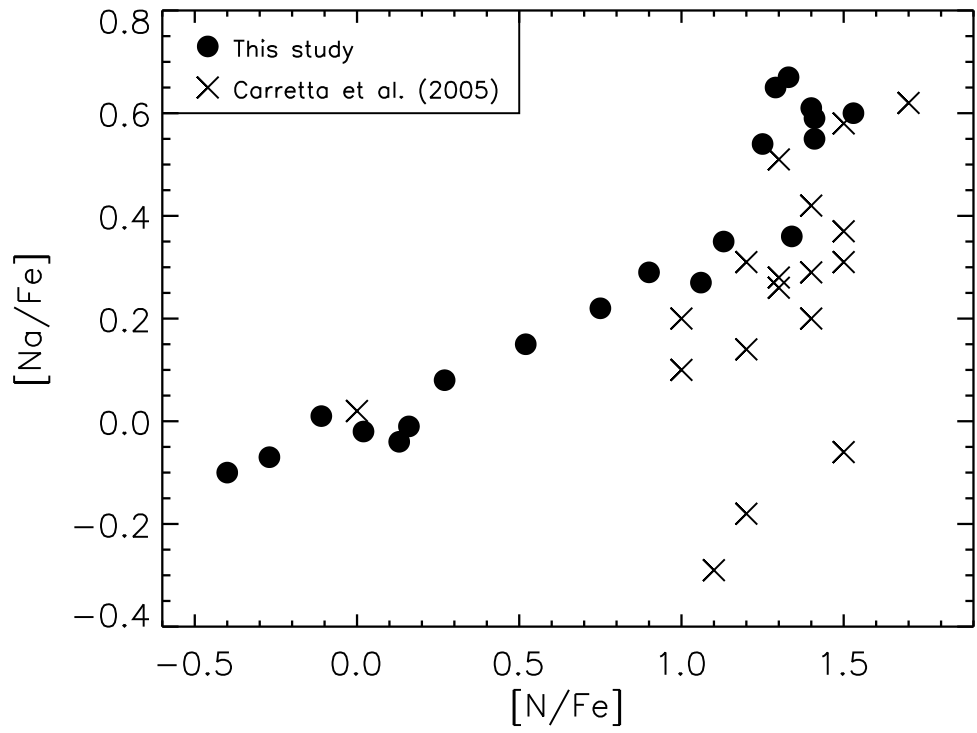}
\caption{[Na/Fe] vs.\ [N/Fe] for this study (closed circles) 
and the \citet{carretta05} sample (crosses). 
\label{fig:nan}}
\end{figure}

For the \citet{carretta05} sample, abundances for other 
light elements known to vary from star to star
have been derived by \citet{gratton01}. For convenience, we consider 
the [Na/Fe] ratio. In Figure \ref{fig:nan} we compare 
[Na/Fe] vs.\ [N/Fe] for our stars and the \citet{carretta05} stars. 
Their minimum value, 
[Na/Fe] = $-$0.29, is slightly lower than our minimum, 
[Na/Fe] = $-$0.10. The maximum [Na/Fe] ratio for \citet{carretta05}, 
[Na/Fe] = +0.62, is comparable to our maximum, 
[Na/Fe] = +0.67. Given that the amplitude of the [Na/Fe] variation in 
\citet{carretta05} exceeds our [Na/Fe] variation, 
we would naively expect the [N/Fe] variation in their sample 
to be comparable to, or larger than, our sample. 
Therefore, it is somewhat unusual that the amplitude of our 
[N/Fe] variation is considerably larger than in 
\citet{carretta05}. Inspection of Figure \ref{fig:nan} shows 
that the [N/Fe] distributions look rather different and we speculate 
that the difference is due in part to difficulties in deriving 
accurate N abundances from CN lines. Although the 
\citet{carretta05} star with [N/Fe] = 0 appears rather unusual compared to 
the bulk of their sample, it lies on the 
distribution defined by our stars. 
One possibility is that measurement uncertainties 
for N and/or Na have affected one or both of the analyses. 
We are not aware of additional analyses of N in large 
numbers of stars in NGC 6752. 

Of the well studied globular clusters, NGC 6752 exhibits one the 
largest amplitudes for light element abundance variations. For other 
well studied globular clusters that also display large abundance 
variations, several have measured N abundances. Work by 
\citet{briley02,briley04} and \citet{cohen02,cohen05b} 
has shown that the 
clusters M5, M13, M15, M71, and 47 Tuc have [N/Fe] ratios that 
cover 2 dex or more. These works are based on very large numbers 
of main sequence stars ($\sim$ 50 - 100), 
albeit using lower resolution spectra with abundances 
determined via the comparison 
of indices measured in observed and synthetic spectra. 
These indices measure the flux removed by molecular features 
relative to nearby continuum bandpasses \citep{briley01}. 
\citet{carretta05} also measured N abundances 
in modest numbers of dwarfs and subgiants 
in the globular clusters NGC 6397 and 47 Tuc. For NGC 6397, their analysis 
of subgiant stars shows that [N/Fe] covers a range of 2.0 dex. For 
47 Tuc, their analysis of dwarfs and subgiants shows a [N/Fe] range of 
1.6 dex. In summary, the amplitude of the N abundance variation in 
this study of NGC 6752 is comparable to that observed in other clusters. 

\subsection{Str{\" o}mgren photometry and N abundances}

\begin{figure}
\epsscale{0.8}
\plotone{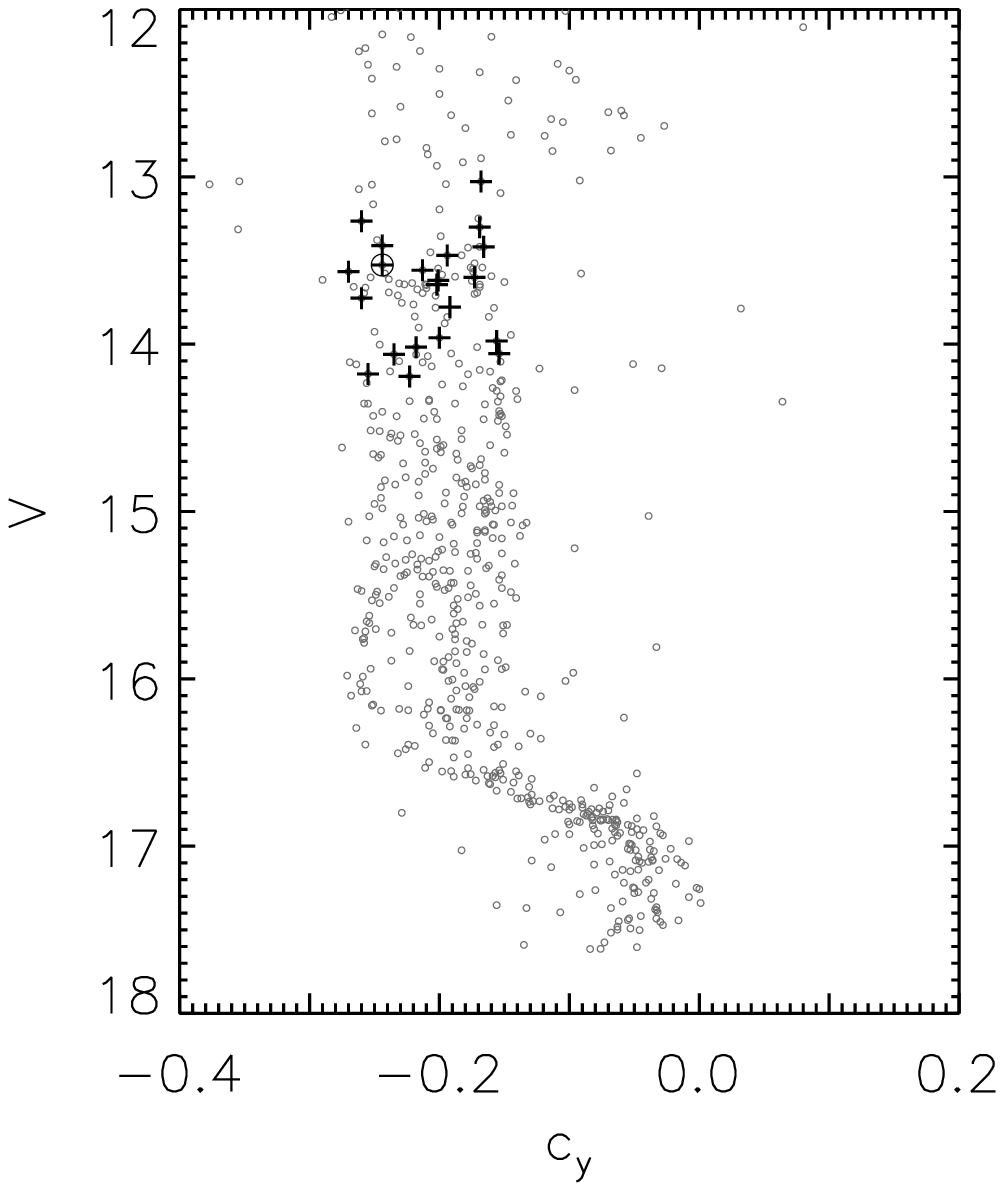}
\caption{$V$ vs.\ $cy$ color-magnitude diagram using 
the \citet{grundahl99} photometry. The plus signs indicate the 
locations of our targets. (The large open circle marks NGC 6752-7.)
\label{fig:cmd2}}
\end{figure}

\begin{figure}
\epsscale{0.8}
\plotone{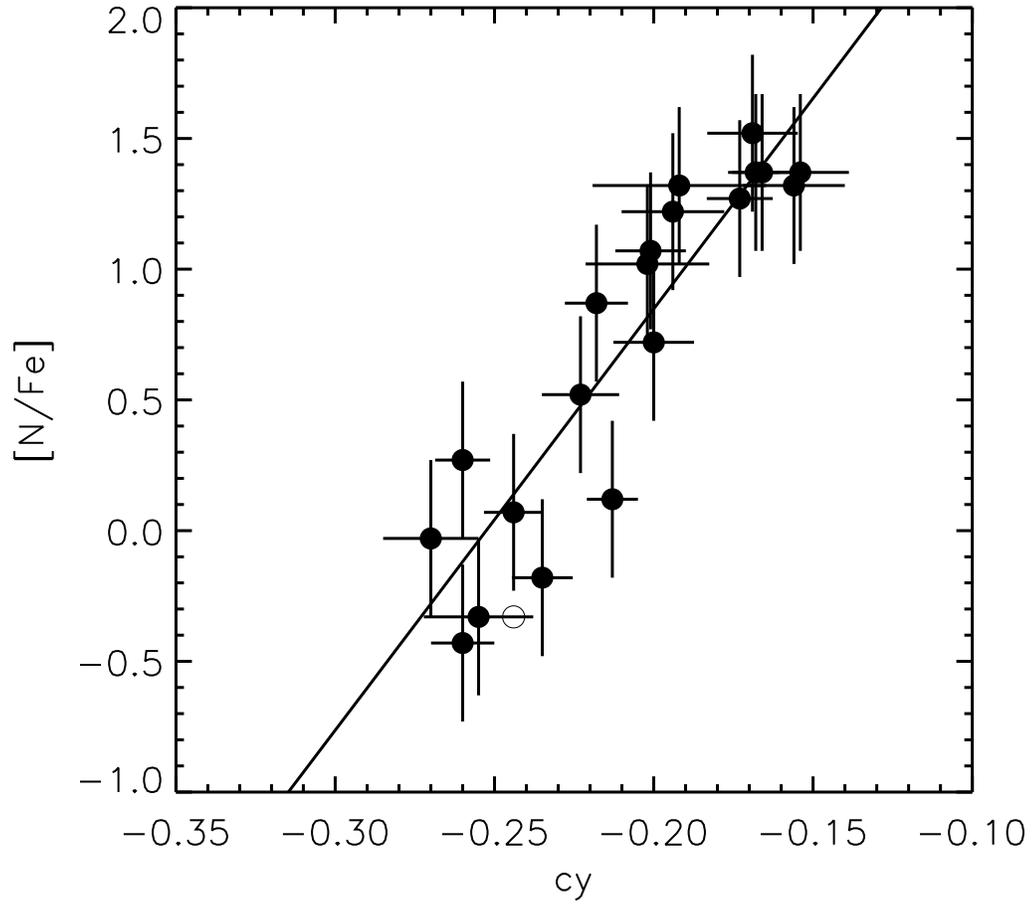}
\caption{[N/Fe] vs.\ $cy$. The straight line is the best fit to the data. 
(The large open circle marks NGC 6752-7.)
\label{fig:ncy}}
\end{figure}

To further investigate the relationship between Str{\" o}mgren 
photometry and N abundances, we introduce a new index $cy$ = $c_1 - (b-y)$. 
This index was identified through simple experiements and is a purely 
empirical index which appears to remove temperature from $c_1$, 
to first order. 
The $cy$ index represents $c_1$ well, but for stars on the lower RGB 
(fainter than the bump) $cy$ appears to show no evolutionary component. 
In Figure \ref{fig:cmd2}, we plot 
$V$ versus $cy$. While there appears to be a large scatter in $cy$ at all 
evolutionary phases, the width and center of the distribution are independent 
of the $V$ magnitude. In Figure \ref{fig:ncy}, 
we plot [N/Fe] versus $cy$. Remarkably, there appears 
to be a linear relation to which we fit 
[N/Fe] = 16.11$\times cy$ + 4.07. The scatter around the fit is only 0.29 dex.  
A linear fit between [N/Fe] and $c_1$ has a scatter of 
0.31 dex and a quadratic fit has a scatter of 0.27 dex. The reduced 
scatter for the quadratic fit confirms the impression from 
Figure \ref{fig:nc1}. However, we prefer the linear relation 
between [N/Fe] and $cy$ which can then be used to explore in more detail  
the distribution of [N/Fe] 
for the large numbers of giant stars for which Str{\" o}mgren 
photometry, but not high-resolution spectra, are available. 

\begin{figure}
\epsscale{0.8}
\plotone{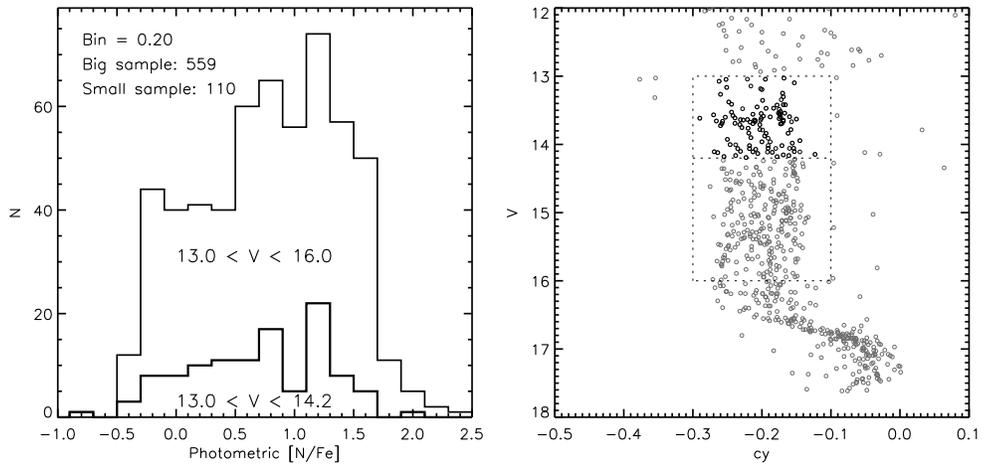}
\caption{The left panel shows the 
[N/Fe] distribution using the relation between [N/Fe] and $cy$. The 
larger histogram shows all stars with 13 $\le$ $V$ $\le$ 16 while the 
smaller histogram 
shows only stars spanned by the calibrating stars. Both sets of 
stars are identified in the $V$ vs.\ $cy$ color-magnitude diagram in the 
right panel. \label{fig:ndist}}
\end{figure}

In Figure \ref{fig:ndist}, we plot the [N/Fe] distribution 
for all giant stars (N = 110) 
with $V$ magnitudes spanned by the calibration stars 
employing the relation given above. 
We find that the distribution increases towards higher [N/Fe] values. 
That is, the distribution is not flat, nor is it bimodal. 
\citet{norris81} found that the CN distribution for NGC 6752 is bimodal 
based on 69 giant stars. 
In the same Figure, we also plot the [N/Fe] distribution for all giant 
stars with magnitudes 13 $\le$ $V$ $\le$ 16 (N = 559). Although it is not clear 
whether the above relation is applicable beyond the magnitudes (i.e., 
stellar parameters) of the calibrating stars, we note that 
both distributions are very similar. These results imply that in 
other clusters, well calibrated $uvby$ photometry could be used
to probe the nitrogen abundance distribution. This may even be possible 
for clusters in the LMC and SMC. 

The $cy$-[N/Fe] relation is valid over the range of parameters covered 
by our sample. Having applied this relation to the entire RGB, we 
recover a similar [N/Fe] distribution as seen within the stars spanned 
by the calibrating stars. Therefore, we speculate that this relation 
may be applicable along the entire RGB. 

\subsection{Correlations between the abundances of nitrogen 
and other elements}

\begin{figure}
\epsscale{0.8}
\plotone{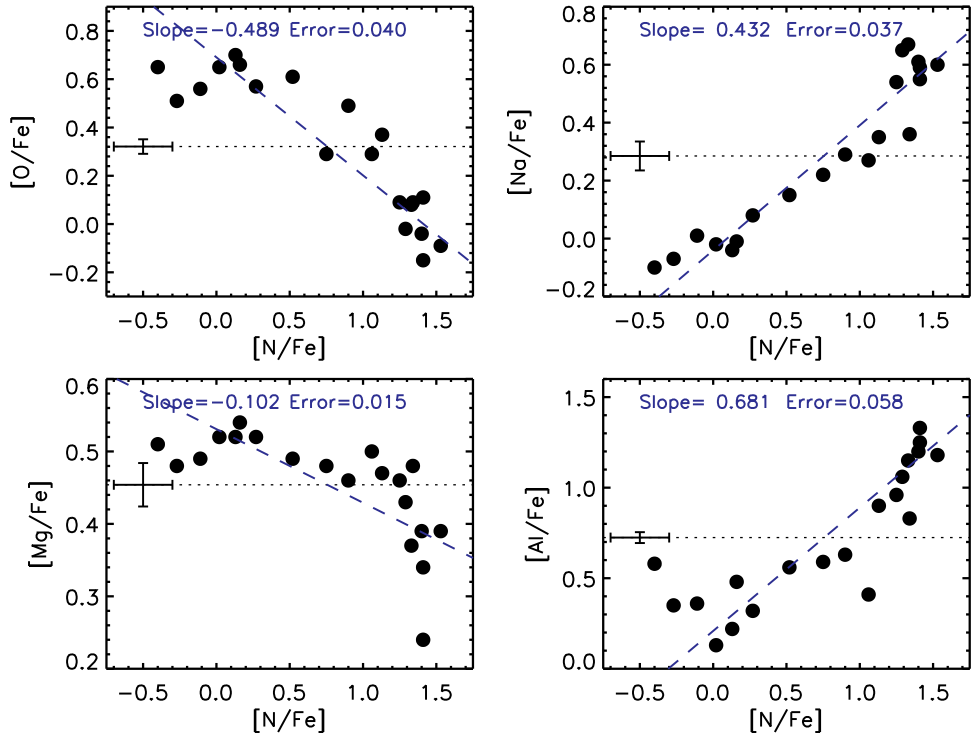}
\caption{[X/Fe] vs.\ [N/Fe] for O, Na, Mg, and Al. The error bar shows 
the 1-$\sigma$ errors (see text for details). 
The dotted line is the mean abundance and the 
dashed line is the linear least squares fit to the data 
(slope and associated error are included). 
\label{fig:ona}}
\end{figure}

\begin{figure}
\epsscale{0.8}
\plotone{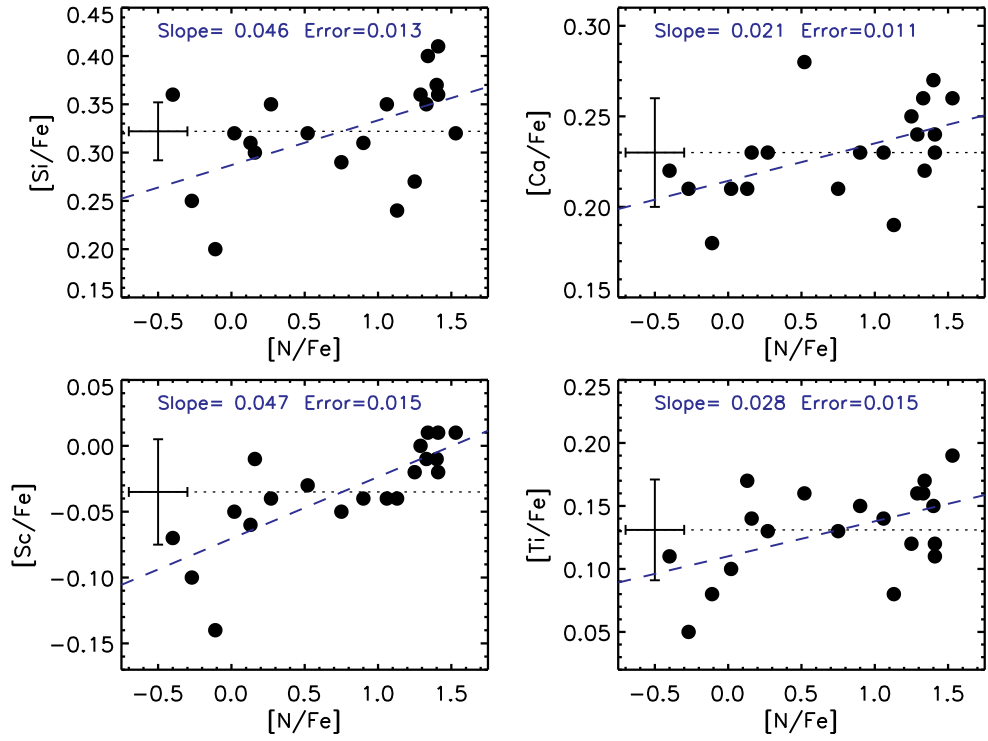}
\caption{Same as Figure \ref{fig:ona} but for Si, Ca, Sc, and Ti.
\label{fig:siti}}
\end{figure}

\begin{figure}
\epsscale{0.8}
\plotone{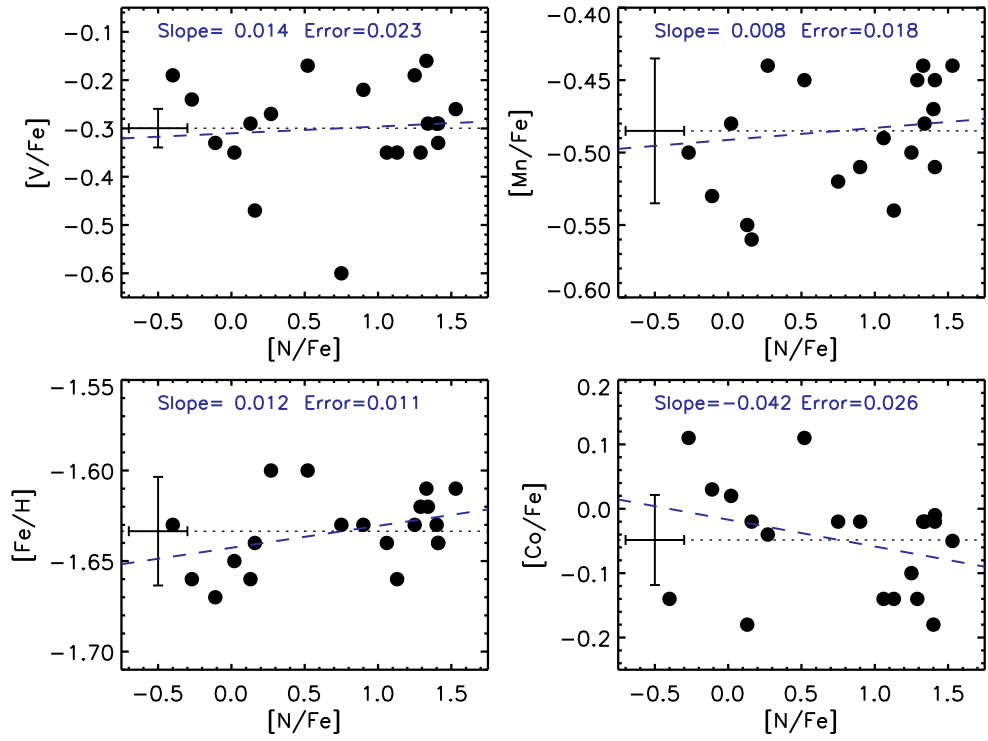}
\caption{Same as Figure \ref{fig:ona} but for V, Mn, Fe, and Co.
\label{fig:vca}}
\end{figure}

\begin{figure}
\epsscale{0.8}
\plotone{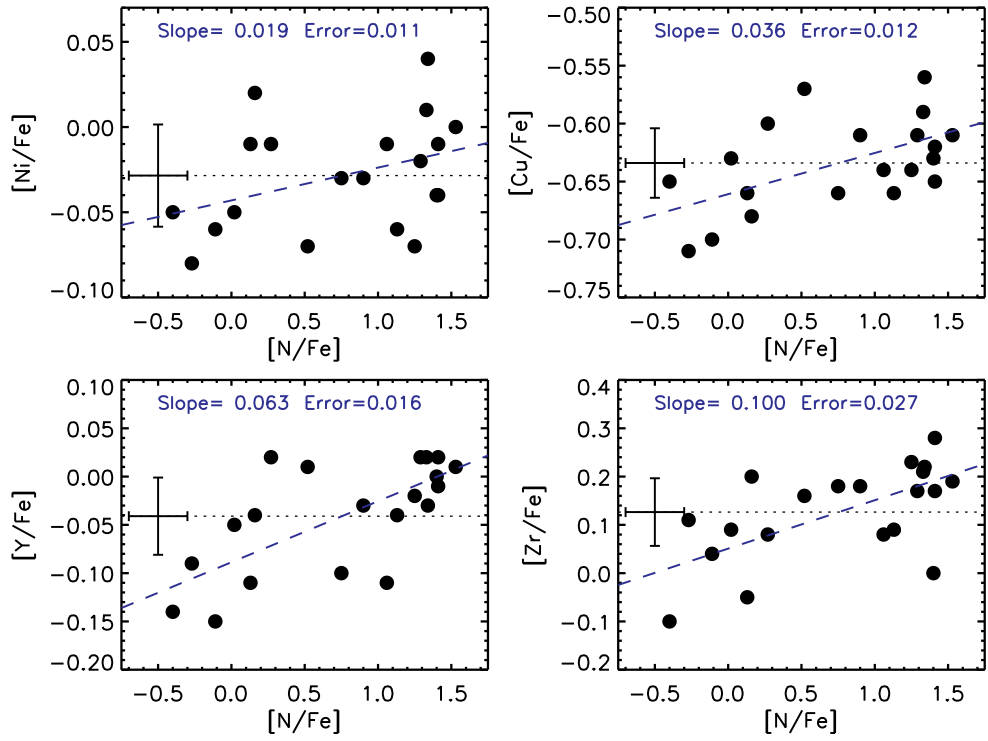}
\caption{Same as Figure \ref{fig:ona} but for Ni, Cu, Y, and Zr.
\label{fig:nizr}}
\end{figure}

\begin{figure}
\epsscale{0.8}
\plotone{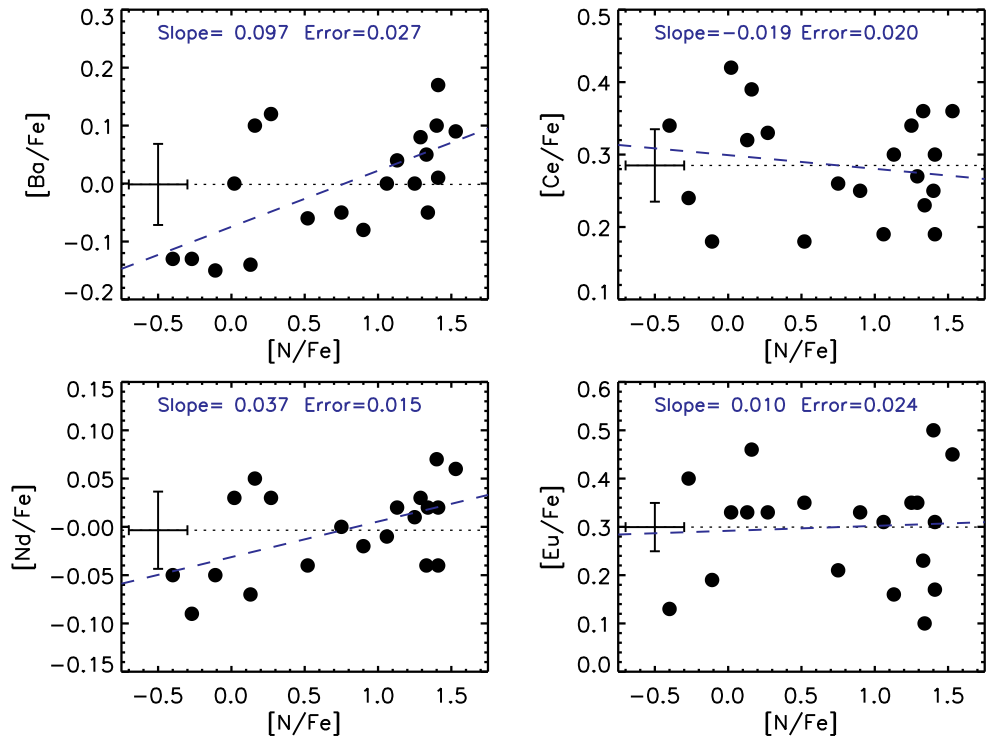}
\caption{Same as Figure \ref{fig:ona} but for Ba, Ce, Nd, and Eu.
\label{fig:baeu}}
\end{figure}

Statistically significant correlations, 
albeit of small amplitude, between Al and the
abundances of Si and heavier elements in NGC 6752's bright 
giant stars (including the RGB bump stars in this analysis) 
were identified by \citet{67522}. Since the N abundances 
exhibit a larger amplitude than the Al abundances, we seek to 
investigate whether there are correlations between N and 
heavier elements. In Figures \ref{fig:ona} to \ref{fig:baeu}, 
we plot the N abundances versus all measured elements in NGC 6752. 
In these Figures, the abundances 
for elements other than N are taken from \citet{grundahl02} and 
\citet{6752,67522}. All studies are homogeneous using the 
same model atmospheres, stellar parameters, and analysis techniques. 

For O, Na, Mg, and Al, the usual correlations with N are evident. 
For completeness, we note that these correlations are significant 
at the 10-$\sigma$ level (i.e., we fit a straight line to the 
data and measure the slope and uncertainty, 
taking into account both the x and y errors).  
We find that the elements Si, Ca, Sc, Ti, Ni, Cu, Y, Zr, Ba, 
and Nd all show correlations with N. 
Although the amplitude of the abundance variations for these heavier elements 
is small ($<$ 0.2 dex), the correlations are statistically significant
at the level 2-$\sigma$ or greater. 
(Even if we increase the x and y errors by 50\%, Si, Sc, Cu, Y, Zr, and Ba 
remain correlated with N at the level 2-$\sigma$ or greater.) 
Taken at face value, 
these correlations indicate that the nucleosynthetic 
source of the N abundance variation must also synthesize small amounts of 
these heavier elements. 
In \citet{67522}, we showed that the abundances are not correlated with 
\teff~and therefore any correlation between N and heavier elements 
is unlikely to be the result of systematic errors in the stellar parameters.

The model atmospheres used in the analysis were 
computed using scaled-solar compositions whereas the stars to which 
these atmospheres were applied do not have scaled-solar compositions. 
Therefore, we were concerned that the correlations between N and 
heavier elements could be due to the use of inappropriate models. 
That is, would the derived abundances change if we used model atmospheres
with the appropriately small or large N, O, Na and Al abundances? 
Or equivalently, does the inclusion of appropriate N, O, Na, and Al 
abundances change the structure of the model atmospheres to such a 
degree that the derived abundances of Si and heavier elements are 
affected? Only small changes in the heavy element abundances 
would be required to remove any correlations with N. 

In Table \ref{tab:err}, we show the abundances for a subset of 
lines derived using four different model atmospheres. 
The model atmospheres and subsequent abundances were computed using 
the MARCS suite of software \citep{marcs75,asplund97}. 
Models 1 and 2 have identical 
\teff, $\log g$, and [Fe/H]. However, Model 1 has a scaled-solar 
composition whereas Model 2 has 
[N/Fe] = +1.5, [O/Fe] = $-$0.1, [Na/Fe] = +0.6, [Mg/Fe] = +0.4, 
and [Al/Fe] = +1.2, values which are essentially identical to those measured
in the N-rich star NGC 6752-2. Similarly Models 3 and 4 have identical 
\teff, $\log g$, and [Fe/H]. However, Model 3 has a scaled-solar 
composition whereas Model 4 has 
[N/Fe] = $-$0.4, [O/Fe] = +0.6, [Na/Fe] = $-$0.1, [Mg/Fe] = +0.5, 
and [Al/Fe] = +0.6, values which are essentially identical to those measured 
in the N-poor star NGC 6752-15. For the purposes of these calculations,  
we assumed a strength of 10m\AA~to 
ensure that the lines are weak and on the linear part of the curve of growth.
A comparison between the abundances 
derived using Models 1 and 2 gauges the uncertainty 
when analyzing N-rich, Na-rich, Al-rich, O-poor, and Mg-poor stars. 
A comparison between the abundances 
derived using Models 3 and 4 gauges the uncertainty for N-poor, Na-poor, 
Al-poor, O-rich, and Mg-rich stars. And a comparison between the abundances 
derived using Models 2 and 4 gauges the uncertainty across the full 
abundance range. Given the small range in \teff~($\sim$ 250K), 
we assume that the calculations would be applicable to all stars in our
sample.  

While the elements considered in Table 2 were chosen somewhat arbitrarily, 
they are neutral and ionized species for which the correlations with N are 
significant (4-$\sigma$ for Y and 3-$\sigma$ for both Si and Zr) and 
for which any correlations, if real, would have considerable impact on our 
understanding of stellar and/or globular cluster evolution. For all
elements, the calculations indicate that adopting model atmospheres
with appropriate compositions would result in only small 
changes to the derived abundances. For the correlations that we 
are investigating, the N-rich stars tend to have higher abundances 
[X/Fe] than their N-poor counterparts. That is, the calculations indicate 
that the correlations between [X/Fe] and [N/Fe] 
would in fact be strengthened had we 
adopted model atmospheres with appropriate compositions. 
Therefore, we tentatively rule out the possibility that the correlations 
between N and heavier elements are due to using model atmospheres with 
scaled-solar compositions. Nevertheless, a more detailed analysis using model 
atmospheres with the appropriate, but presently unknown, abundances of He 
and C would be of great interest. 

As far as we are aware, correlations between the abundances of 
light elements (N and Al) and elements heavier than 
Si have not been noted in any other globular clusters. 
We attribute our detection primarily to the highly accurate and 
homogeneous stellar parameters derived from the \citet{grundahl99} 
Str{\" o}mgren photometry as well as to the high quality spectra 
and large sample size. 

\subsection{Implications for globular cluster and stellar evolution}

\begin{figure}
\epsscale{0.8}
\plotone{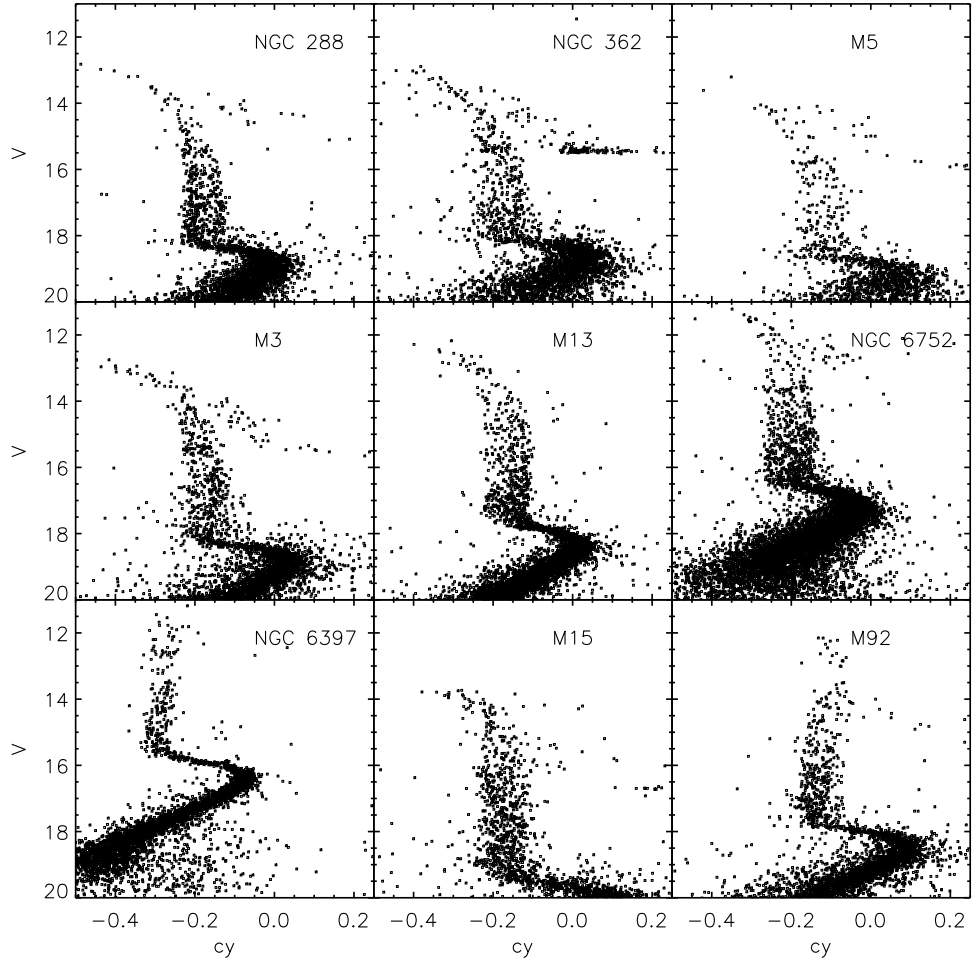}
\caption{The $V$ vs.\ $cy$ color-magnitude diagrams for a 
sample of clusters taken from \citet{grundahl99}. All clusters 
exhibit a large range in $cy$ at all evolutionary stages. 
\label{fig:cyall}}
\end{figure}

Having demonstrated that the Str{\" o}mgren $c_1$ index (or $cy$ index) 
directly traces 
the N abundance, we confirm and reiterate the claim by \citet{grundahl00}
that every globular cluster exhibits a considerable star-to-star 
variation in N abundances at all evolutionary stages as seen via 
the $c_1$ index. In Figure \ref{fig:cyall}
we plot $V$ versus $cy$ for 
a subset of the \citet{grundahl99} clusters. 
Every cluster displays a dispersion in $cy$ comparable to that 
seen in NGC 6752 indicating that all globular clusters exhibit a 
$\simeq$ 2.0 dex dispersion in [N/Fe], the value found in NGC 6752. 
This Figure clearly shows that 
all clusters exhibit a large dispersion in $cy$, i.e., 
N abundances, at all evolutionary phases. Note that these clusters 
cover a range in metallicities: 
[Fe/H] $\simeq$ $-$1.2 (NGC 288, NGC 362, M5),
[Fe/H] $\simeq$ $-$1.6 (M3, M13, NGC 6752), and 
[Fe/H] $\simeq$ $-$2.0 (NGC 6397, M15, M92). 

Pollution by ejecta from intermediate-mass AGB stars and/or 
massive stars is the currently favored explanation for the 
star-to-star abundance variation of light elements. 
Both scenarios are problematic, with issues relating to 
the initial mass function and the predicted nucleosynthesis yields 
\citep{fenner04,decressin06}. 

The handful of clusters for which N abundances have been measured in 
large samples of stars show a variety of N abundance distributions. 
M71 and 47 Tuc appear bimodal whereas M5, M13, and M15 do not appear bimodal 
(\citealt{cohen05b} and references therein). As discussed above, the 
[N/Fe] distribution for NGC 6752, using the relation between 
N and Str{\" o}mgren photometry, appears to increase towards increasing 
[N/Fe]. Such a distribution is similar to that seen in the 
comparably metal-poor cluster M13 (and possibly also in M5). 
Curiously, both NGC 6752 and M13 exhibit 
bimodal CN distributions, as do all clusters in which CN can be 
measured \citep{smith87}. 
The varying [N/Fe] distributions in clusters of different 
metallicities will provide important 
observational constraints upon the origin of the abundance anomalies 
in future globular cluster chemical evolution studies 
similar to that conducted by \citet{fenner04}. 

The correlations between N and heavy elements presented in this paper 
offer support for contributions from 
both AGB and massive stars to the observed abundance anomalies. 
Although quantitative yields of 
$s$-process elements from intermediate-mass metal-poor 
AGB stars are rare, 
$s$-process 
elements are expected to be synthesized, to some degree, in these stars 
\citep{busso01}. 
Fe-peak and $\alpha$ elements 
are synthesized by massive stars, at all metallicities \citep{chieffi04}. 
Given that not every element is 
correlated with N at a statistically significant level, and that the 
amplitude of the heavy element abundance variation is small, our 
results do not offer definitive support for either the 
massive star or AGB pollution scenarios. 

As mentioned, target selection via the Str{\" o}mgren $c_1$ index 
has allowed us to sample the full range of the 
star-to-star abundance variations, at this luminosity. Of great interest 
would be a further study of the full range of the abundance variations 
at different luminosities. An analysis of the most N-poor and N-rich stars, 
as selected via the $c_1$ index, would truly gauge any variation in the 
abundance anomalies as a function of evolutionary status. That the 
abundance variations for O-Al have been found in unevolved stars suggests 
that primordial pollution processes rather than internal evolutionary 
processes must be the dominant mechanism, but these 
observations do not entirely preclude 
an internal evolutionary mechanism being a second-order effect. 
Indeed, the brightest stars in M13 appear to exhibit a 
systematic abundance variation for O, Na, Mg, and Al 
with luminosity \citep{sneden04a}. 
Such trends,
if observed in additional clusters, 
would present new challenges for stellar evolution 
and stellar nucleosynthesis. 

The \citet{grundahl99} 
Str{\" o}mgren photometry extends to additional globular clusters. 
It would be of great interest to obtain high-resolution, high S/N spectra 
for a large sample of stars within a narrow luminosity range 
in other globular clusters in order to 
see whether the correlations between light and heavy elements are also 
present. The key is to again use carefully calibrated 
Str{\" o}mgren photometry to obtain homogeneous stellar parameters, 
such as in \citet{grundahl02}. 

Star-to-star helium abundance variations are expected when the 
C-Al abundance anomalies are synthesized during hydrogen burning. 
NGC 2808 \citep{dantona04} and 
$\omega$ Cen \citep{norris04} show evidence for He variations, based 
primarily on the interpretation of color-magnitude diagrams. 
Stellar models and isochrones computed by \citet{salaris06} 
show that the location of the RGB bump (weakly) depends upon the 
He abundance. Recently, \citet{carretta07} found new evidence for 
He variations in NGC 6218 (M12) based on RGB stars. 
They looked at the luminosity functions 
for a sample of Na-rich and a sample of Na-poor stars and found that 
the RGB bump differed by 0.05 mag, corresponding to a difference 
in $Y$ of 0.05 between the Na-poor and Na-rich populations. 
Although the correlation is not significant, we find 
marginal evidence for an increase in [Fe/H] with increasing [N/Fe]. 
Such a correlation may be the signature of possible He variations. 

Finally, one of the most stringent observational 
constraints upon the origin of the abundance anomalies comes from 
C+N+O, which is constant to within a factor of 2 \citep{smith05}. 
Predicted yields 
from AGB models are unable to satisfy this constraint 
\citep{lattanzio06,karakas06}. 
The ideal check of the constancy of C+N+O 
would be to measure C from CH, N from NH, and O from [OI] using 
high-resolution spectra for a sample that covers the full 
range of the abundance variations. We have measured 2 of these 3 indicators 
but unfortunately our spectra do not cover the 4300\AA~CH 
molecular lines. Measurement of C from CH for a subset of 
the most N-rich and N-poor stars in this sample is highly desired 
to examine the constancy of C+N+O and to search for correlations 
between C+N+O versus N.  

\section{Concluding remarks}
\label{sec:summary}

In this paper we present measurements of N abundances in 21 bright 
giant stars near the RGB bump of the globular cluster NGC 6752. 
The sample was chosen to span the full range of the Str{\" o}mgren $c_1$ 
index at this luminosity. 
The amplitude of the N abundance variation at the sample's luminosity 
is 1.95 dex. We confirm that the N abundances are 
correlated with the $c_1$ index. We find a linear relation between 
the N abundances and a new index $cy$ = $c_1$ $- (b-y)$. We apply 
this new relation to large numbers of 
RGB stars and find that the [N/Fe] distribution 
increases towards higher values. Broader implications are that 
all globular clusters show large star-to-star variations in 
their $cy$ indices (i.e., N abundances) at all evolutionary stages. 

We find that the N abundances 
are correlated with the light elements O, Na, Mg, and Al, a feature 
that is seen in every well studied globular cluster. However, we find 
for the first time that the N abundances are also correlated with 
the abundances of Si and heavier elements. While such correlations are 
statistically significant, the amplitude of the variation is small
($<$ 0.2 dex variation in [X/Fe] as [N/Fe] varies by 2 dex). We 
attribute the detection to the large sample size, high quality data, 
and most importantly to the homogeneous and precise stellar 
parameters obtained from the \citet{grundahl99} Str{\" o}mgren photometry. 

Analysis using model atmospheres with the appropriate N, O, Na, and Al 
abundances gives very similar abundances to those 
derived using model atmospheres with scaled-solar compositions. 
Therefore, the correlations between N and heavier elements are unlikely 
to be due to using models with scaled-solar compositions. In fact, for 
the subset of elements investigated (Si, Y, and Zr), the correlations with N 
would be even more significant had we used models with appropriate 
compositions. 

Of great interest would be to search for such correlations in an 
additional sample of stars at a different luminosity within this 
cluster and in other clusters. If correlations 
between light and heavy elements are again 
identified, they would offer support to both the AGB and 
massive star pollution scenarios for explaining the star-to-star 
abundance variation of light elements. 

\acknowledgments

This research has made use of the SIMBAD database,
operated at CDS, Strasbourg, France and
NASA's Astrophysics Data System. DY thanks Mike Bessell, 
Amanda Karakas, John Norris for helpful discussions. We thank the 
anonymous referee for helpful comments. This research was 
supported in part by NASA through the American Astronomical Society's Small 
Research Grant Program.

\clearpage

\clearpage

\clearpage

\clearpage

\clearpage

\clearpage

\clearpage

\clearpage

\clearpage

\clearpage

\clearpage

\clearpage

\clearpage

\clearpage

\begin{deluxetable}{llcccccccr} 
\tabletypesize{\footnotesize}
\tablecolumns{10} 
\tablewidth{0pc} 
\tablecaption{Stellar parameters and nitrogen abundances.\label{tab:param}}
\tablehead{ 
\colhead{Name 1\tablenotemark{a}} &
\colhead{Name 2} &
\colhead{RA (J2000)} &
\colhead{DEC (J2000)} &
\colhead{$V$} &
\colhead{\teff} &
\colhead{log $g$} &
\colhead{$\xi_t$} &
\colhead{[Fe/H]} &
\colhead{[N/Fe]}
}
\startdata
\ldots & NGC6752-0  & 19:11:03 & $-$59:59:32 & 13.03 & 4699 & 1.83 & 1.47 & $-$1.62 & 1.37 \\
B2882  & NGC6752-1  & 19:10:47 & $-$60:00:43 & 13.27 & 4749 & 1.95 & 1.41 & $-$1.58 & 0.27 \\
B1635  & NGC6752-2  & 19:11:11 & $-$60:00:17 & 13.30 & 4779 & 1.98 & 1.39 & $-$1.59 & 1.52 \\
B2271  & NGC6752-3  & 19:11:00 & $-$59:56:40 & 13.41 & 4796 & 2.03 & 1.42 & $-$1.64 & 0.07 \\
B611   & NGC6752-4  & 19:11:33 & $-$60:00:02 & 13.42 & 4806 & 2.04 & 1.40 & $-$1.61 & 1.37 \\
B3490  & NGC6752-6  & 19:10:34 & $-$59:59:55 & 13.47 & 4804 & 2.06 & 1.40 & $-$1.61 & 1.22 \\
B2438  & NGC6752-7  & 19:10:57 & $-$60:00:41 & 13.53 & 4829 & 2.10 & 1.33 & $-$1.84 & $-$0.33 \\
B3103  & NGC6752-8  & 19:10:45 & $-$59:58:18 & 13.56 & 4910 & 2.15 & 1.33 & $-$1.62 & 0.12 \\
B3880  & NGC6752-9  & 19:10:26 & $-$59:59:05 & 13.57 & 4824 & 2.11 & 1.38 & $-$1.63 & $-$0.03 \\
B1330  & NGC6752-10 & 19:11:18 & $-$59:59:42 & 13.60 & 4836 & 2.13 & 1.37 & $-$1.60 & 1.27 \\
B2728  & NGC6752-11 & 19:10:50 & $-$60:02:25 & 13.62 & 4829 & 2.13 & 1.32 & $-$1.64 & 1.07 \\
B4216  & NGC6752-12 & 19:10:20 & $-$60:00:30 & 13.64 & 4841 & 2.15 & 1.34 & $-$1.62 & 1.02 \\
B2782  & NGC6752-15 & 19:10:49 & $-$60:01:55 & 13.73 & 4850 & 2.19 & 1.35 & $-$1.61 & $-$0.43 \\
B4446  & NGC6752-16 & 19:10:15 & $-$59:59:14 & 13.78 & 4906 & 2.24 & 1.32 & $-$1.60 & 1.32 \\
B1113  & NGC6752-19 & 19:11:23 & $-$59:59:40 & 13.96 & 4928 & 2.32 & 1.29 & $-$1.61 & 0.72 \\
\ldots & NGC6752-20 & 19:10:36 & $-$59:56:08 & 13.98 & 4929 & 2.33 & 1.32 & $-$1.59 & 1.32 \\
\ldots & NGC6752-21 & 19:11:13 & $-$60:02:30 & 14.02 & 4904 & 2.33 & 1.29 & $-$1.61 & 0.87 \\
B1668  & NGC6752-23 & 19:11:12 & $-$59:58:29 & 14.06 & 4916 & 2.35 & 1.27 & $-$1.62 & 1.37 \\
\ldots & NGC6752-24 & 19:10:44 & $-$59:59:41 & 14.06 & 4948 & 2.37 & 1.15 & $-$1.65 & $-$0.18 \\
\ldots & NGC6752-29 & 19:10:17 & $-$60:01:00 & 14.18 & 4950 & 2.42 & 1.26 & $-$1.64 & $-$0.33 \\
\ldots & NGC6752-30 & 19:10:39 & $-$59:59:47 & 14.19 & 4943 & 2.42 & 1.27 & $-$1.62 & 0.52 \\
\enddata

\tablenotetext{a}{Star names from \citet{buonanno86}.}

\end{deluxetable}

\begin{deluxetable}{ccccccccc} 
\tabletypesize{\footnotesize}
\tablecolumns{9} 
\tablewidth{0pc} 
\tablecaption{Derived abundances for model atmospheres with scaled-solar and 
non-scaled-solar compositions.\label{tab:err}}
\tablehead{ 
\colhead{Wavelength} &
\colhead{Species} &
\colhead{eV} &
\colhead{$\log gf$} &
\colhead{EW} &
\multicolumn{4}{c}{Abundance: log$\epsilon$(X)} \\
\colhead{} &
\colhead{} &
\colhead{} &
\colhead{} &
\colhead{} &
\colhead{Model 1\tablenotemark{a}} &
\colhead{Model 2\tablenotemark{b}} &
\colhead{Model 3\tablenotemark{c}} &
\colhead{Model 4\tablenotemark{d}}
}
\startdata
5690.43 & 14.0 & 4.93 & $-$1.83 & 10.0 & 5.98 & 6.00 & 6.02 & 6.02 \\
6270.22 & 26.0 & 2.86 & $-$2.51 & 10.0 & 5.33 & 5.32 & 5.40 & 5.38 \\
6271.28 & 26.0 & 3.33 & $-$2.76 & 10.0 & 6.13 & 6.13 & 6.20 & 6.18 \\
6416.92 & 26.1 & 3.89 & $-$2.74 & 10.0 & 5.64 & 5.70 & 5.70 & 5.75 \\
5509.91 & 39.1 & 0.99 & $-$1.01 & 10.0 & 0.09 & 0.18 & 0.19 & 0.25 \\
6134.55 & 40.0 & 0.00 & $-$1.28 & 10.0 & 2.12 & 2.10 & 2.23 & 2.18 \\
\enddata

\tablenotetext{a}{\teff = 4780K, $\log g$ = 2.0, [Fe/H] = $-1.60$, scaled solar compositions}
\tablenotetext{b}{Same as Model 1 but with [N/Fe] = +1.5, [O/Fe] = $-$0.1, [Na/Fe] = +0.6, [Mg/Fe] = +0.4, [Al/Fe] = +1.2}
\tablenotetext{c}{\teff = 4850K, $\log g$ = 2.2, [Fe/H] = $-1.60$, scaled solar compositions}
\tablenotetext{d}{Same as Model 3 but with [N/Fe] = $-$0.4, [O/Fe] = +0.6, [Na/Fe] = $-$0.1, [Mg/Fe] = +0.5, [Al/Fe] = +0.6}

\end{deluxetable}

\end{document}